\documentclass[MSNbibl,nameyear,seceqn,dvips]{arxstspdf}
\usepackage{dcolumn}
\usepackage{flushend}
\usepackage{stfloats}
\volume{29}
\issue{1}
\pubyear{2014}
\firstpage{2}
\lastpage{8}
\doi{10.1214/12-STS413} %kopijuoti is PTS
\makeatletter
\newcolumntype{d}[1]{D{.}{.}{#1}}
\def\var{\operatorname{var}}
\def\Normal{\operatorname{Normal}}
\def\tY{\widetilde{Y}}
\def\tW{\widetilde{W}}
\def\tU{\widetilde{U}}
\def\tepsilon{\widetilde{\varepsilon}}
\def\density{\mathrm{density}}

\setattribute{abstract}   {width}  {345pt}
\setattribute{keyword}    {width}  {345pt}

\makeatother

\begin{document}
\begin{frontmatter}

\title{Estimating the Distribution of Dietary Consumption Patterns\thanksref{t1}}
\relateddois{t1}{Discussed in \relateddoi{d}{10.1214/13-STS448} and \relateddoi{d}{10.1214/14-STS466}.}
\runtitle{Dietary Consumption Patterns}
% kai straipsnis turi susijusiu diskusiju ir rejoinder'iu
%rejoinder at \relateddoi{r}{10.1214/00-STSXXXX}.}

\begin{aug}
\author{\fnms{Raymond J.} \snm{Carroll}\ead[label=e10]{carroll@stat.tamu.edu}}
\runauthor{R. J. Carroll}

\affiliation{Texas A\&M University}

\address{Raymond J. Carroll is Distinguished Professor of Statistics,
Nutrition and Toxicology,
Department of Statistics, Texas A\&M University, 3143 TAMU, College
Station, Texas 77843-3143,
USA \printead{e10}.}

\end{aug}

% ABSTRACT
%
\begin{abstract}
In the United States the preferred method of obtaining dietary intake
data is the 24-hour dietary recall, yet the measure of most interest is
usual or long-term average daily intake, which is impossible to
measure. Thus, usual dietary intake is assessed with considerable
measurement error. We were interested in estimating the population
distribution of the Healthy Eating Index-2005 (HEI-2005), a
multi-component dietary quality index involving ratios of interrelated
dietary components to energy, among children aged 2--8 in the United
States, using a national survey and incorporating survey weights. We
developed a highly nonlinear, multivariate zero-inflated data model
with measurement error to address this question. Standard nonlinear
mixed model software such as SAS NLMIXED cannot handle this problem. We
found that taking a Bayesian approach, and using MCMC, resolved the
computational issues and doing so enabled us to provide a realistic
distribution estimate for the HEI-2005 total score. While our
computation and thinking in solving this problem was Bayesian, we
relied on the well-known close relationship between Bayesian posterior
means and maximum likelihood, the latter not computationally feasible,
and thus were able to develop standard errors using balanced repeated
replication, a survey-sampling approach.
\end{abstract}

% KEYWORDS

\end{frontmatter}

%s1 #&#
\section{Introduction}\label{intro}

We (Zhang et al., \citeyear{Zhaetal11N2}, which has many additional references)
confronted the following problem in dietary assessment. A summary of
the key issues follows:

\begin{table*}
\caption{Description of the HEI-2005 scoring system. Except for saturated fat
and SoFAAS (calories from solid fats, alcoholic beverages and added
sugars), density is obtained by multiplying usual intake by 1000 and
dividing by usual intake of kilo-calories. For saturated fat, density
is $9 \times100 $ usual saturated fat (grams) divided by usual
calories, that is, the percentage of usual calories coming from usual
saturated fat intake. For SoFAAS, the density is the percentage of
usual intake that comes from usual intake of calories, that is, the
division of usual intake of SoFAAS by usual intake of calories. Here,
``DOL'' is dark green and orange vegetables and legumes. The total
HEI-2005 score is the sum of the individual component scores}\label{tab:heidescription}
\begin{tabular*}{\textwidth}{@{\extracolsep{\fill}}lcl@{}}
\hline
\textbf{Component} & \textbf{Units} & \textbf{HEI-2005 score calculation} \\
\hline
Total fruit & cups & $\min(5, 5\times(\density/0.8))$ \\
Whole fruit & cups & $\min(5, 5\times(\density/0.4))$\\
Total vegetables & cups & $\min(5, 5\times(\density/1.1))$\\
DOL & cups & $\min(5, 5\times(\density/0.4))$\\
Total grains & ounces & $\min(5, 5\times(\density/3))$\\
Whole grains & ounces & $\min(5, 5\times(\density/1.5))$\\
Milk & cups & $\min(10, 10\times(\density/1.3))$\\
Meat and beans & ounces & $\min(10, 10\times(\density/2.5))$\\
Oil & grams & $\min(10, 10\times(\density/12))$\\
Saturated fat & \% of energy& if $\density\ge15$ score${} = 0$\\
&  & else if $\density\le7$ score${}= 10$\\
& & else if $\density>10$ score $= 8- (8\times(\density- 10)/5)$\\
& & else, score${} = 10 - (2\times(\density- 7)/3)$\\
Sodium & milligrams & if $\density\ge2000$ score${}=0$\\
& & else if $\density\le700$ score${}=10$\\
& & else if $\density\ge1100$ \\
& & \quad score $ = 8 - \{8\times( \density- 1100)/(2000 -
1100)\}$\\
& & else score $ = 10 - \{2\times(\density- 700)/(1100 - 700)\}$\\
SoFAAS & \% of energy& if $\density\ge50$ score${}= 0$\\
&  & else if $\density\le20$ score${}=20$\\
& & else score $= 20 - \{20\times(\density- 20)/(50 - 20)\}$\\
\hline
\end{tabular*}
\end{table*}

\begin{itemize}
\item Nutritional surveys conducted in the United States typically use
24-hour (24hr) dietary recalls to obtain intake data, that is, an
assessment of what was consumed in the past 24 hours.
\item Because dietary recommendations are intended to be met over time,
nutritionists are interested in ``usual'' or long-term average daily intake.
\item Dietary intake is assessed with considerable measurement error. A
very large part of the measurement error is that diet has great
within-person variability, so that a snapshot of two days of recall
cannot hope to capture an individual's average intake over a year.
There are also other sources of error besides the fact that diets vary
greatly across different days. In the 24hr recall instruments used, the
instrument uses a ``multi-pass'' approach that circles around to try to
elicit better memory of what was eaten. The method is actually quite
good in getting people to remember what they ate, but errors arise
through estimation of portion size, which can be both too large and too small.
\item Consumption patterns of dietary components vary widely; some are
consumed daily by almost everyone, while others are episodically
consumed so that their 24-hour recall data are zero-inflated. Further,
these components are correlated with one another.
\item Nutritionists are interested in dietary components collectively
to capture patterns of \textit{usual} dietary intake, and thus need
multivariate models for usual intake.
\item We knew of no standard frequentist software that had any hope of
fitting the model and obtaining answers.
\end{itemize}

%t1 #&#

One way to capture dietary patterns is by scores. The Healthy Eating
Index-2005 (HEI-2005) is a scoring system based on a priori knowledge
of dietary recommendations and is on a scale of 0 to 100. See Table~\ref{tab:heidescription} for a list of these components and the
standards for scoring, and see Guenther et al. (\citeyear{Gueetal08}) and \citet{GueReeKre08} for
details. Ideally, it consists of the \textit{usual} intake of 6
episodically consumed and thus 24hr-zero inflated foods, 6
daily-consumed dietary components, adjusts these for energy (caloric)
intake, and gives a score to each component. The total score is the sum
of the individual component scores. Higher scores indicate greater
compliance with dietary guidelines and, therefore, a healthier diet.
The questions we addressed were to estimate the distribution of the
HEI-2005 total score and to estimate the \% of American children who
are eating an alarmingly poor diet, defined by a total score less than 40.

To answer public health questions such as these that can have policy
implications, we (Zhang et al., \citeyear{Zhaetal11N2}) built a novel multivariate
measurement error model for estimating the distributions of usual
intakes, one that accounts for measurement error and multivariate
zero-inflation, and had a special covariance structure associated with
the zero-inflation. Previous attempts to fit even simple versions of
this model, using nonlinear mixed effects software, failed because of
the complexity and dimensionality of the model. We used survey-weighted
Monte Carlo computations to fit the model with uncertainty estimation
coming from balanced repeated replication. The methodology was
illustrated using the HEI-2005 to assess the diets of children aged 2--8
in the United States. This work represented the first analysis of joint
distributions of usual intakes for multiple food groups and nutrients.

The 12 HEI-2005 components represent 6 episodically consumed food
groups (total fruit, whole fruit, total vegetables, dark green and
orange vegetables and legumes or DOL, whole grains and milk), 3
daily-consumed food groups (total grains, meat and beans, and oils),
and 3 other daily-consumed dietary components (saturated fat; sodium;
and calories from solid fats, alcoholic beverages and added sugars, or
SoFAAS). The crucial statistical aspect of the data is that six of the
food groups are zero-inflated. The percentages of reported
nonconsumption of total fruit, whole fruit, whole grains, total
vegetables, DOL and milk on any single day are 17\%, 40\%, 42\%, 3\%,
50\% and 12\%, respectively.

%s2 #&#
\section{Data and the HEI-2005 Scores}\label{sec:sec2}

We are interested in the usual intake of foods for children aged 2--8.
The data available to us came from the National Health and Nutrition
Examination Survey, 2001--2004 (NHANES). The data consisted of $n
=2638$ children, each of whom had a survey weight $w_i$ for
$i=1,\ldots,n$. In addition, one or two 24hr dietary recalls were
available for each individual. Along with the dietary variables, there
are covariates such as age, gender, ethnicity, family income and dummy
variables that indicate a weekday or a weekend day, and whether the
recall was the first or second reported for that individual.

Using the 24hr recall data reported, for each of the episodically
consumed food groups, two variables are defined: (a) whether a food
from that group was consumed; and (b) the amount of the food that was
reported on the 24hr recall. For the 6 daily-consumed food groups and
nutrients, only one variable indicating the consumption amount is
defined. In addition, the amount of energy that is calculated from the
24hr recall is of interest. The number of dietary variables for each
24hr recall is thus $12+6+1 = 19$. The observed data are $Y_{ijk}$ for
the $i$th person, the $j$th variable and the $k$th replicate,
$j=1,\ldots,19$ and $k=1,\ldots,m_i$. In the data set, at most two
24hr recalls were observed, so that $m_i \leq2$. Set $\tY_{ik} =
(Y_{i1k},\ldots,Y_{i,19,k})^{\mathrm{T}}$, where:
\begin{itemize}
\item$Y_{i, 2\ell-1, k} = {}$Indicator of whether dietary component
\#$\ell$ is consumed, with $\ell= 1, 2, 3, 4, 5, 6$.
\item$Y_{i, 2\ell,k} = {}$Amount of food \#$\ell$ consumed. This
equals zero, of course, if none of food \#$\ell$ is consumed,
with $\ell= 1, 2, 3, 4, 5, 6$.
\item$Y_{i,\ell+6,k} ={}$Amount of nonepisodically consumed food or
nutrient \#$\ell$,
with $\ell= 7, 8, 9, 10, 11, 12$.
\item$Y_{i,19,k} ={}$Amount of energy consumed as reported by the 24hr recall.
\end{itemize}

%s3 #&#
\section{Model and Methods}\label{sec:sec3}
%s3.1 #&#
\subsection{Basic Model Description} \label{sec:sec2_1}

Observed data will be denoted as $Y$, and covariates in the model will
be denoted as $X$. As is usual in measurement error problems, there
will also be latent variables, denoted by $W$.

We used a probit threshold model. Each of the 6 episodically consumed
foods has 2 sets of latent variables, one for consumption and one for
amount, while the 6 daily-consumed foods and nutrients as well as
energy have 1 latent variable each, for a total of 19. The latent
random variables are $\varepsilon_{ijk}$ and $U_{ij}$, where $(U_{i1},
\ldots, U_{i,19}) = \Normal(0,\Sigma_u)$ and $(\varepsilon
_{i1k},\ldots,\varepsilon_{i,19,k}) = \Normal(0,\Sigma_{\varepsilon})$
are mutually independent. In this model, food $\ell= 1,\ldots,6$ being
consumed on day $k$ is equivalent to observing the binary $Y_{i,2\ell
-1, k}$, where
%
%e3.1 #&#
\begin{eqnarray}
\label{eq:qkipnis03}\qquad&& Y_{i,2\ell-1, k} = 1\quad\Longleftrightarrow\nonumber \\
&&  W_{i,2\ell-1, k}\\
 &&\quad= X_{i,2\ell-1,k}^{\mathrm{T}}\beta_{2\ell-1} +
U_{i,2\ell-1} + \varepsilon_{i,2\ell-1, k} > 0.\nonumber
\end{eqnarray}
If the food is consumed, we model the amount reported, $Y_{i, 2\ell,
k}$, as
%
%e3.2 #&#
\begin{eqnarray}
\label{eq:qkipnis04} &&\bigl[g_{\mathrm{tr}}(Y_{i,2\ell, k},
\lambda_\ell) \vert Y_{i,2\ell-1,
k}=1 \bigr] \nonumber\\
&&\quad= W_{i,2\ell, k}
\\
&&\quad= X_{i,2\ell,k}^{\mathrm{T}}\beta_{2\ell} +
U_{i,2\ell} + \varepsilon _{i,2\ell, k},\nonumber
\end{eqnarray}
where $g_{\mathrm{tr}}(y,\lambda) = \sqrt{2} \{g(y,\lambda) - \mu
(\lambda)\}/\sigma(\lambda)$, $g(y,\lambda)$ is the usual Box--Cox
transformation with transformation parameter $\lambda$, and $\{\mu
(\lambda),\sigma(\lambda)\}$ are the sample mean and standard
deviation of $g(y,\lambda)$, computed from the nonzero food data. This
standardization is a convenient device to improve the numerical
performance of our algorithm without affecting our conclusions.

The reported consumption of daily-consumed foods or nutrients $\ell=
7, \ldots, 12$ is modeled as
%
%e3.3 #&#
\begin{eqnarray}
\label{eq:qkipnis05a} &&g_{\mathrm{tr}}(Y_{i, \ell+6, k},\lambda_\ell)\nonumber\\
&&\quad=
W_{i,\ell+6, k} \\
&&\quad= X_{i,\ell+6,k}^{\mathrm{T}}\beta_{\ell+6} +
U_{i, \ell+6} + \varepsilon _{i,\ell+6, k}.\nonumber
\end{eqnarray}
Finally, energy is modeled as
%
%e3.4 #&#
\begin{eqnarray}\label{eq:qkipnis05b}
\qquad g_{\mathrm{tr}}(Y_{i, 19, k},\lambda_{13})&=& W_{i,19, k}
\nonumber
\\[-8pt]
\\[-8pt]
\nonumber& =&
X_{i,19,k}^{\mathrm{T}}\beta_{19} + U_{i, 19} +
\varepsilon_{i,19, k}.
\end{eqnarray}
As seen in (\ref{eq:qkipnis04})--(\ref{eq:qkipnis05b}), different
transformations $(\lambda_1,\break \ldots,\lambda_{13})$ are allowed to be used
for the different types of dietary components.

In summary, there are latent variables $\tW_{ik} =\break  (W_{i1k},\ldots,W_{i,
19, k})^{\mathrm{T}}$, latent random effects\vspace*{1pt} $\tU_i =\break
(U_{i1},\ldots,U_{i,19})^{\mathrm{T}}$, fixed effects $(\beta
_1,\ldots,\beta
_{19})$, and design matrices $(X_{i1k},\ldots,X_{i,19,k})$. Define
$\tepsilon_{ik} = (\varepsilon_{i1k},\break \ldots,\varepsilon_{i, 19, k})^{\mathrm{T}}$.
For\vspace*{1pt} mutually independent random variables $\tU_i = \Normal(0,\Sigma
_u)$ and $\tepsilon_{ik} = \Normal(0,\Sigma_{\varepsilon})$, the
latent variable model is
%
%e3.5 #&#
\begin{equation}
W_{ijk} = X_{ijk}^{\mathrm{T}}\beta_j +
U_{ij} + \varepsilon_{ijk}. \label{eq:qe01}
\end{equation}

%s3.2 #&#
\subsection{Restriction on the Covariance Matrix}\label{secrest}
Two necessary restrictions are set on $\Sigma_{\varepsilon}$. First,
following Kipnis et al. (\citeyear{Kipetal09}), $\varepsilon_{i,2\ell-1, k}$ and
$\varepsilon_{i, 2\ell, k}$ ($\ell=1,\ldots,6$) are set to be
independent. Second, in order to technically identify $\beta_{2\ell
-1}$ and the distribution of $U_{i, 2\ell-1}$ ($\ell=1,\ldots,6$),
we require that $\var(\varepsilon_{i, 2\ell-1, k}) = 1$, because
otherwise the marginal probability of consumption of component $\#\ell
$ is $\Phi\{(X_{i,2\ell-1,k}^{\mathrm{T}}\beta_{2\ell-1} + U_{i,
2\ell
-1})/\var^{1/2}(\varepsilon_{i, 2\ell-1, k})\}$, and thus components of
$\beta$ and $\Sigma_u$ would be identified only up to the scale $\var
^{1/2}(\varepsilon_{i, 2\ell-1, k})$.

It is easiest to see the problem in the case of two episodically
consumed dietary components and energy. In this case,
%
%e3.6 #&#
\begin{equation}
\label{eq:qszmc01} \Sigma_{\varepsilon} = \pmatrix{ 1 & 0 & s_{13} &
s_{14} & s_{15}
\cr
0 & s_{22} & s_{23} &
s_{24} & s_{25}
\cr
s_{13} & s_{23} & 1 &
0 & s_{35}
\cr
s_{14} & s_{24} & 0 &
s_{44} & s_{45}
\cr
s_{15} & s_{25} &
s_{35} & s_{45} & s_{55}
\cr
}.
\end{equation}
The difficulty with parameterizations of (\ref{eq:qszmc01}) is that
the cells that are not constrained to be $0$ or $1$ cannot be left
unconstrained, otherwise (\ref{eq:qszmc01}) need not be a covariance
matrix, that is, positive semidefinite. Zhang et al. (\citeyear{Zhaetal11N2}) developed
an unconstrained parameterization that results in the structure (\ref
{eq:qszmc01}).

%s3.3 #&#
\subsection{The Use of Sampling Weights}\label{weights}

We used the survey sample weights from NHANES both in the model fitting
procedure and, after having fit the model, in estimating the
distributions of usual intake. As a referee pointed out, the role of
sampling weights in Bayesian analyses is controversial. In our problem,
most of the variables that were used to construct the sampling weights
were in our model, and we thus expected that a weighted and unweighted
analysis would lead to very similar parameter estimates (posterior
means) for the model. This was indeed the case. However, the sampling
weights definitely are needed in estimating the population distribution
that is representative of the US population, not just the sample.
Having fit the model, estimation of distribution was done in a routine
frequentist manner using the survey weights. We assessed standard
errors by Balanced Repeated Replication (BRR).

%t2 #&#
\begin{table*}[b]
\tabcolsep=0pt
\caption{Estimated distributions of the usual
intake HEI-2005 scores. Standard errors are given in Zhang et al.
(\citeyear{Zhaetal11N2}). The total score is the sum of the individual scores. Here,
``DOL'' is dark green and orange vegetables and legumes. Also,
``SoFAAS'' is calories from solid fats, alcoholic beverages and added
sugars}\label{tab:tab3}
\begin{tabular*}{\textwidth}{@{\extracolsep{4in minus 4in}}ld{2.2}d{2.2}d{2.2}d{2.2}d{2.2}d{2.2}d{2.2}d{2.2}@{}}
\hline
& & \multicolumn{7}{c@{}}{\textbf{Percentile}}\\%[-6pt]
\ccline{3-9}
%& & \multicolumn{7}{c@{}}{\hrulefill}\\
\textbf{Component} & \multicolumn{1}{c}{\textbf{Mean}} & \multicolumn{1}{c}{$\mathbf{5th}$} &
\multicolumn{1}{c}{$\mathbf{10th}$} & \multicolumn{1}{c}{$\mathbf{25th}$} &
\multicolumn{1}{c}{$\mathbf{50th}$} & \multicolumn{1}{c}{$\mathbf{75th}$} &
\multicolumn{1}{c}{$\mathbf{90th}$} & \multicolumn{1}{c@{}}{$\mathbf{95th}$}\\
\hline
Total fruit &3.55&0.87&1.31&2.33&3.90&5.00&5.00&5.00\\
Whole fruit &3.14&0.49&0.82&1.71&3.24&5.00&5.00&5.00\\
Total vegetables &2.16&1.02&1.24&1.63&2.10&2.62&3.15&3.48\\
DOL &0.62&0.05&0.09&0.21&0.45&0.86&1.38&1.76\\
Total grains &4.81&3.92&4.23&4.79&5.00&5.00&5.00&5.00\\
Whole grains &0.90&0.16&0.24&0.43&0.75&1.21&1.74&2.13\\
Milk &6.77&2.15&2.96&4.62&6.91&9.67&10.00&10.00\\
Meat and beans &7.22&4.23&4.83&5.91&7.21&8.64&10.00&10.00\\
Oil &5.92&3.37&3.83&4.69&5.77&7.01&8.25&9.07\\
Saturated fat &5.16&0.00&1.09&3.18&5.38&7.48&8.53&8.96\\
Sodium &4.52&1.25&2.05&3.31&4.62&5.83&6.85&7.44\\
SoFAAS &8.73&2.15&3.60&6.02&8.73&11.42&13.81&15.21\\[3pt]
Total score &53.50&37.42&40.74&46.73&53.68&60.36&65.87&68.96\\
\hline
\end{tabular*}
\end{table*}

%s3.4 #&#
\subsection{Distribution of Usual Intake and the HEI-2005
Scores}\label{subsec:dhei}

Zhang et al. (\citeyear{Zhaetal11N2}) use MCMC to estimate $\Sigma_u$, $\Sigma
_{\varepsilon}$ and $\beta_j$ for $j=1,\ldots,19$. From that, it is
straightforward to estimate the distribution of usual intakes and the
usual HEI-2005 component and total scores. To see how this works,
consider the first episodically consumed dietary component, a food
group. Following Kipnis et al. (\citeyear{Kipetal09}), we define the usual intake for
an individual on the weekend to be the expectation of the reported
intake conditional on the person's random effects $\tU_i$. Let the
$(q,p)$ element of $\Sigma_{\varepsilon}$ be denoted as $\Sigma
_{\varepsilon,q,p}$. As in Kipnis et al., define\looseness=1
\[
g_{\mathrm{tr}}^*\{v,\lambda,\Sigma_{\varepsilon,q,p}\} = g_{\mathrm{tr}}^{-1}(v,
\lambda) + \frac{1}{2}\Sigma_{\varepsilon,q,p} \frac
{\partial^2 g_{\mathrm{tr}}^{-1}(v,\lambda)}{\partial v^2}.
\]\looseness=0
Then, following the convention of Kipnis et al. (\citeyear{Kipetal09}), the person's
usual intake of the first episodically consumed dietary component is
defined as
\[
T_{i1} = \Phi\bigl(X_{i1}^{\mathrm{T}}\beta_1 +
U_{i1}\bigr)g_{\mathrm{tr}}^* \bigl(X_{i2}^{\mathrm{T}}
\beta_2 + U_{i2},\lambda_1,\Sigma_{\varepsilon
,2,2}
\bigr).
\]
Usual intake for the other episodically consumed food groups is defined
similarly, and similarly for the daily-consumed components. With such
definitions it is straightforward to generate, via Monte-Carlo, a
survey-weighted estimate of the population distribution of usual intake.

%s4 #&#
\section{Why a Bayesian Approach to Estimation}\label{sec:howtoestimate}

Our model (\ref{eq:qkipnis04})--(\ref{eq:qkipnis05b}) is a highly
nonlinear, mixed effects model with many latent variables and nonlinear
restrictions on the covariance matrix $\Sigma_{\varepsilon}$. As
discussed in Section \ref{subsec:dhei}, we can estimate relevant
distributions of usual intake in the population if we can estimate
$\Sigma_u$, $\Sigma_{\varepsilon}$ and $\beta_j$ for $j=1,\ldots,19$. We
have found that working within a Bayesian paradigm is a convenient way
to do this computation and thus solve the problem. We used this
approach because standard software such as NLMIXED simply could not
handle the problem, while thinking computationally in a Bayesian way
using MCMC was straightforward.
Indeed, Zhang et al. (\citeyear{Zhaetal11N1}) have shown that even considering a single
food group plus energy is a challenge for the NLMIXED procedure, both
in time and in convergence, and using this method for the entire
HEI-2005 constellation of dietary components is impossible.

Kipnis et al. (\citeyear{Kipetal09}) were able to get estimates of parameters
separately for each food group using the nonlinear mixed effects
program NLMIXED in SAS with sampling weights. While this gives
estimates of $\beta_j$ for $j=1,\ldots,19$, it only gives us parts of the
covariance matrices $\Sigma_u$ and $\Sigma_{\varepsilon}$, and not all
the entries. Using the 2001--2004 NHANES data, we have verified that our
estimates and the subset of the parameters that can be estimated by one
food group at a time using NLMIXED are in close agreement, and that
estimates of the distributions of usual intake and HEI-2005 component
scores are also in close agreement.

Full technical details of the MCMC model fitting procedure are given in
Zhang et al. (\citeyear{Zhaetal11N2}).

%s5 #&#
\section{Empirical Work} \label{sec:sec5}

%s5.1 #&#
\subsection{Basic Analysis}\label{sec:sec5.1}

As stated previously, we analyzed data from the 2001--2004 National
Health and Nutrition Examination Survey (NHANES) for children ages 2--8.
We used the dietary intake data to calculate the 12 HEI-2005 components
plus energy. In addition, besides age, gender, race and interaction
terms, two covariates were employed, along with an intercept. The first
was a dummy variable indicating whether or not the recall was for a
weekend day (Friday, Saturday or Sunday) because food intakes are known
to differ systematically on weekends and weekdays. The second was a
dummy variable indicating whether the 24hr recall was the first or
second such recall, the idea being that there may be systematic
differences attributable to the repeated administration of the instrument.

%s5.2 #&#
\subsection{Estimation of the HEI-2005 Scores}\label{sec:5.3}

Table \ref{tab:tab3} presents the first estimates of the distribution
of HEI-2005 scores for a vulnerable subgroup of the population, namely,
children aged 2--8 years. A~previous analysis of 2003--04 NHANES data,
looking separately at 2--5 year olds and 6--11 year olds, was limited to
estimates of mean usual HEI-2005 scores [59.6 and 54.7, resp.;
see Fungwe et al. (\citeyear{Funetal09})]. The mean scores noted here are comparable
to those and reinforce the notion that children's diets, on average,
are far from ideal. However, this analysis provides a more complete
picture of the state of US children's diets. By including the scores at
various percentiles, we estimate that only 5\% of children have a score
of 69 or greater and another 10\% have scores of 41 or lower. While not
in the Table, we also estimate that the $99$th percentile is $74$.
This analysis suggests that virtually all children in the US have
suboptimal diets and that a sizeable fraction (10\%) have alarmingly
low scores (41 or lower).\looseness=1

We have also considered whether our multivariate model fitting
procedure gives reasonable marginal answers. To check this, we note
that it is possible to use the SAS procedure NLMIXED \textit{separately
for each component} to fit a model with one episodically consumed food
group or daily-consumed dietary component together with energy. The
marginal distributions of each such component done separately are quite
close to what we have reported in Table \ref{tab:tab3}, as is our
mean, which is $53.50$ compared to the mean of $53.25$ based on
analyzing one HEI-2005 component at a time with the NLMIXED procedure.
The only case where there is a mild discrepancy is in the estimated
variability of the energy-adjusted usual intake of oils, likely caused
by the NLMIXED procedure itself, which has an estimated variance $9$
times greater than our estimated variance.

Of course, it is the distribution of the HEI-2005 total score that
cannot be estimated by analysis of one component at a time.

Finally, we also estimated the distribution of the total score as
developed by a single 24hr, and completely ignoring the difference
between a 24hr and usual intake. A single 24HR estimated that nearly
30\% of children have an alarmingly poor diet (total score $\leq40$)
versus the 10\% we think is realistic. This difference is enormous. If
the 30\% figure were to be believed, which we do not think it should
be, this could have major policy implications.

%s5.3 #&#
\subsection{Computing and Data}\label{canddata}
Our programs were written in Matlab. The programs, along with the
NHANES data we used, are available in the \textit{Annals of Applied Statistics} online archive
associated with Zhang et al. (\citeyear{Zhaetal11N2}). Because of the public health
importance of the problem, the National Cancer Institute has contracted
for the creation of a SAS program that performs our analysis. It will
allow any number of episodically and daily-consumed dietary components.
The first draft of this program, written independently in a different
programming language, gives almost identical results to what we have
obtained, verifying that our results are not the product of a
programming error.\looseness=1

%s6 #&#
\section{Discussion}\label{sec:sec7}

There are many important questions in dietary assessment that have not
been able to be answered because of a lack of multivariate models for
complex, zero-inflated data with measurement errors and especially a
lack of ability to fit such multivariate models. Nutrients and foods
are not consumed in isolation, but rather as part of a broader pattern
of eating. There is reason to believe that these various dietary
components interact with one another in their effect on health,
sometimes working synergistically and sometimes in opposition.
Nonetheless, simply characterizing various patterns of eating has
presented an enormous statistical challenge. Until now, descriptive
statistics on the HEI-2005 have been limited to examination of either
the mean total score or only a single energy-adjusted component at a
time, neither of them relevant for the distribution of the total score.
This has precluded characterization of various patterns of dietary
quality as well as any subsequent analyses of how such patterns might
relate to health.

The Bayesian methodology presented in Zhang et al. (\citeyear{Zhaetal11N2}) presents a
workable solution to these problems which has already proven valuable.
In May 2010, just as we were submitting the paper, a White House Task
Force on Childhood Obesity created a report. They had wanted to set a
goal of all children having a total HEI score of 80 or more by 2030,
but when they learned we estimated only 10\% of the children ages 2--8
had a score of 66 or higher, they decided to set a more realistic
target. The facility to estimate distributions of the multiple
component scores simultaneously will be important in tracking progress
toward that goal.

In some respects, our model is complex, but then the problem is
complex. There are simple solutions for some subproblems of the
HEI-2005,\vadjust{\goodbreak} for example, estimating its mean. However, our problem is a
setting where the actual distribution is of interest, not just the
mean. Recognizing that the simplest approach, ignoring measurement
error entirely, led to unrealistic estimates of the percentage of
children with alarmingly poor diets (Section \ref{sec:5.3}), we
realized that a measurement error analysis was required. We worked on
this problem for nearly 2 years before realizing that the only
practical way forward was to take a Bayesian approach to computation.
Estimating distributions of usual (i.e., measurement error corrected)
dietary patterns is important for public health and policy. We are
extending our methods to the newly announced HEI-2010 score, which has
more dietary components, and preliminary results are encouraging.

\section*{Acknowledgments}

Carroll's research was supported by a grant from the National Cancer
Institute (R27-CA057030). He thanks the co-authors of Zhang et al.
(\citeyear{Zhaetal11N2}) for their work on the project. This publication is based in
part on work supported by Award Number KUS-CI-016-04, made by King
Abdullah University of Science and Technology (KAUST), and also in part
by the Spanish Ministry of Science and Innovation (project MTM
2011-22664 which is co-funded by FEDER).\vfill


%suskaldyti doi

% imsref loaded by akundreckaite, 2013-04-02 12:02:02
%

\end{document}